\documentstyle[prd,aps,epsfig]{revtex}
\newcommand{\dd}{{\rm d}}

\newcommand{\DD}{{\cal D}}

\begin{document}
\draft
\title{The stress-energy tensor for trans-Planckian cosmology}
\author{Martin Lemoine$^1$, Musongela Lubo$^2$,
J\'er\^ome Martin$^1$ and Jean--Philippe
Uzan$^{3,1}$\vskip0.5cm}
\address{(1) Institut d'Astrophysique de Paris, CNRS,
             98bis Bd. Arago, F--75014 Paris (France)\\
         \vskip0.25cm
         (2) M\'ecanique et Gravitation, Universit\'e de Mons--Hainaut,\\
             6 avenue du Champ de Mars, B--7000 Mons (Belgium)\\
         \vskip0.25cm
         (3) Laboratoire de Physique Th\'eorique, CNRS-UMR 8627,
             B\^at. 210,\\ Universit\'e Paris XI,
             F--91405 Orsay Cedex (France).}
\date{\today}
\maketitle
\begin{abstract}
This article presents the derivation of the stress-energy tensor of a
free scalar field with a general non-linear dispersion relation in
curved spacetime. This dispersion relation is used as a phenomelogical
description of the short distance structure of spacetime following the
conventional approach of trans-Planckian modes in black hole physics
and in cosmology.  This stress-energy tensor is then used to discuss
both the equation of state of trans-Planckian modes in cosmology and
the magnitude of their backreaction during inflation. It is shown that
gravitational waves of trans-Planckian momenta but subhorizon
frequencies cannot account for the form of cosmic vacuum energy
density observed at present, contrary to a recent claim. The
backreaction effects during inflation are confirmed to be important
and generic for those dispersion relations that are liable to induce
changes in the power spectrum of metric fluctuations. Finally, it is
shown that in pure de Sitter inflation there is no modification of the
power spectrum except for a possible magnification of its overall
amplitude independently of the dispersion relation.
\end{abstract}
\pacs{PACS numbers: 98.80.Cq}
\maketitle
\section{Introduction}\label{sec_intro}

The inflationary paradigm provides an appealing framework to describe
the very early phase of the evolution of the Universe, notably because
it produces in a natural way the seeds necessary to the formation of the
cosmological large scale structures~\cite{inf}. These initial density
fluctuations are generated during inflation with an almost scale
invariant power spectrum through the parametric amplification of the
quantum fluctuations of the inflaton scalar field, with the
accelerated Friedman-Lema\^{\i}tre scale factor acting as a classical
pump field. This can also be seen as particle production due to the
breakdown of adiabaticity in the evolution of the quantum modes of the
field as the wavelengths are stretched beyond the horizon.

 In spite of their successes, most models of inflation are subject to
the {\it trans-Planckian problem}, namely the phase of accelerated
expansion lasts sufficiently long that cosmological length scales
today corresponded to scales much smaller than the Planck scale at the
beginning of inflation~\cite{mb,bm}. Depending on one's point of view,
this can be seen either as a problem, i.e.  the celebrated predictions
depend on unknown trans-Planckian physics, or as a blessing, since
inflation then opens a window on physics beyond the Planck
scale. Several studies have recently tackled the issue of the
robustness of inflation to changes in super-Planck physics by adopting
a phenomenological approach initially developed in the context of
black hole physics, where a very similar problem arises~\cite{un,cj},
and in close relation to analogous problems in condensed matter
physics~\cite{volovik}. It consists in modifying the standard
dispersion relation of a free scalar field for wavelengths smaller
than the Planck length and in calculating the resulting power spectrum
of inflation produced metric fluctuations~\cite{mb,bm,niemeyer,KW}.
This approach is further motivated by the fact that the evolution of
the scalar and tensor modes of metric fluctuations can be adequately
described by free scalar fields propagating in a background
spacetime~\cite{mukhanov92}. It should be noted that this is only a
phenomenological approach to the problem of trans-Planckian physics,
and that it is not unique. For instance, modifications of the
canonical quantum operator commutation relations inspired from string
theories~\cite{kempf} have also been shown to affect the power
spectrum of metric fluctuations~\cite{gree1,KN01,EGK01}.

 Interestingly, concrete examples of dispersion relations leading to
modifications of the power spectrum with respect to the standard
(linear dispersion relation) predictions have been
exhibited~\cite{mb,bm} but only in the case where the dispersion
relation becomes complex. Even though complex dispersion relations are
ordinary in classical physics or in quantum mechanics, they represent
a problematic situation in the context of quantum field theory. It has
also been shown that dispersion relations such that the evolution of
the quantum mode is adiabatic all throughout the inflationary phase up
to horizon crossing cannot lead to significant modifications of the
power spectrum~\cite{np}. With respect to the above example of
dispersion relations, indeed adiabaticity is broken at the point where
the physical frequency vanishes and the dispersion relation becomes
complex. However it has been argued that such dispersion relations
which break adiabaticity would lead to a possibly severe backreaction
problem, i.e. the energy density contained in the modes would become
greater than the background energy
density~\cite{tanaka,starobinsky}. Therefore modifications to the
power spectrum may arise if adiabaticity is broken, but at the expense
of the creation of a possibly large amount of energy
density. Unfortunately this amount could not be quantified rigorously
as the authors did not have at their disposal the stress-energy tensor
of a theory with modified dispersion relation.

 In this context, our main objective is to present a rigorous
derivation of the stress-energy tensor of a free scalar field with
non-linear dispersion relation. Such a dispersion relation breaks
local Lorentz invariance as it implies the existence of a preferred
reference frame, however it is crucial to maintain general covariance
in order to achieve consistency with the Einstein equations (and
notably the conservation of the stress-energy tensor). This problem
has been examined in a number of studies, where an effective general
covariant Lagrangian with explicit Lorentz invariance breaking has
been constructed with the introduction of a dynamical unit timelike
vector field whose role is to define the preferred rest
frame~\cite{aether,jacobson}.  An effective Lagrangian describing a
free scalar theory with a quartic dispersion relation could then be
constructed along these lines, and the stress-energy tensor has been
calculated for the particular case of the Corley-Jacobson dispersion
relation~\cite{jacobson,jacobson2}. In the present paper we present a
non-trivial extension of this latter study to the general case in
which the squared frequency is a general analytic function of the
squared momentum. Note that in Friedman-Lema\^{\i}tre-Robertson-Walker
(FLRW) cosmology, a preferred rest frame exists and coincides with the
homogeneous isotropic spatial sections. Finally one should point out
that the method given below cannot be used to derive the stress-energy
tensor for theories with unmodified dispersion relations but modified
canonical commutation relations as considered in
Refs.~\cite{kempf,gree1,KN01,EGK01}.

  We then use this stress-energy tensor to address two points raised
recently in the literature. We first discuss the claim~\cite{mbk}
according to which a bath of gravitons of super-Planck momenta but
frequencies much smaller than the Hubble expansion rate (hence a
particular non-linear dispersion relation), can explain the form of
vacuum energy seen in the Universe today. In order to do so, we calculate
the energy density, the pressure and the equation of state of these
quanta; we show that these gravitons possess neither the correct
energy density nor the correct equation of state.  As a second
application we discuss the issue of backreaction of trans-Planckian
modes in inflationary cosmology, using the expression for the energy
density contained in trans-Planckian modes. 

 This article is organized as follows. In Section~\ref{sec_2} we
formulate a covariant Lagrangian (\S~\ref{sec_2a}) including extra
terms to provide a modified dispersion relations and calculate the
corresponding stress-energy tensor
(\S~\ref{sec_3}). Then, we specify our result to the case of 
cosmology (\S~\ref{sec_2c}). Section~\ref{sec_4} discusses the two main
applications of our results, namely the equation of state of the
trans-Planckian modes (\S~\ref{subsec_4.1}) and the backreaction
problem (\S~\ref{subsec_4.2} and \S~\ref{subsec_4.3}).  Our results
are summarized in Section~\ref{sec_concl}. For the sake of clarity, we
gathered notations and derivations of various identities used in the
calculation of the stress-energy tensor in Appendix~\ref{AppA};
Appendix~\ref{app_B} presents the detailed derivation of the
stress-energy tensor in the simpler case of the Corley-Jacobson
dispersion relation. We use natural units in which $\hbar=k_{\rm
B}=c=1$, and the metric $g_{\mu\nu}$ carries signature $(-,+,+,+)$.

\section{Covariant Lagrangian and stress--energy tensor}\label{sec_2}

In this section, we first introduce a general covariant formulation of
a Lagrangian describing a free scalar field with modified dispersion
relation following the procedure described by Jacobson and
Mattingly~\cite{jacobson,jacobson2}, and derive the corresponding
energy-momentum tensor. We try to remain as general as possible with
respect to the background spacetime ${\cal M}$ and its metric, and
defer the detailed study of a FLRW spacetime to Section~\ref{sec_2c}.

\subsection{Definitions and covariant Lagrangian}\label{sec_2a}

The action for a free scalar field with modified dispersion relation
takes the form~\cite{jacobson,jacobson2}
\begin{equation}\label{action}
S_{_\phi}=\int{\rm d}^4x\sqrt{-g}({\cal L}_{_\phi}+{\cal L}_{_{\rm
cor}}+{\cal L}_u),
\end{equation}
where ${\cal L}_{_\phi}$ is the standard Lagrangian of a minimally
coupled free scalar field
\begin{equation}
{\cal L}_{_\phi}=-\frac{1}{2}g^{\mu\nu}\partial_\mu
\phi\partial_\nu\phi.
\end{equation}
Two contributions have been added to this Lagrangian in order to
introduce the modified dispersion relation. As already mentioned in
the introduction, a modified dispersion relation breaks local Lorentz
invariance. In such a situation, a covariant formulation of the
corresponding theory can be carried out by introducing a unit timelike
vector field $u^{\mu }$ defining a preferred rest
frame~\cite{jacobson,jacobson2}.  In Eq.~(\ref{action}), the first
term ${\cal L}_{_{\rm cor}}$ is responsible for the non-linear part of
the dispersion relation and ${\cal L}_u$ describes the dynamics of the
vector field $u^\mu$. These two corrective Lagrangians take the form
\begin{equation}\label{lcor}
{\cal L}_{_{\rm cor}}=- \sum_{n,p\leq n} b_{np}
\left(\DD^{2n}\phi\right)\left(\DD^{2p}\phi\right),
\quad 
{\cal L}_u=-\lambda(g^{\mu\nu}u_\mu u_\nu +1)-
d_1F^{\mu \nu}F_{\mu \nu}.
\end{equation}
The tensor $F_{\mu \nu }$ is defined by $F_{\mu \nu }\equiv \nabla
_{\mu }\,u_{\nu}-\nabla_\nu\,u_\mu$, where $\nabla_{\mu }$ is the
covariant derivative associated with the metric $g_{\mu \nu}$;
$\lambda$ is a Lagrange multiplier and the coefficients $b_{np}$ and
$d_1$ are arbitrary. The derivative operators ${\cal D}^{2n}$ are
defined further below. The overall Lagrangian maintains general
covariance, which will notably ensure the conservation of the
stress-energy tensor. The value of the Lagrange multiplier $\lambda$
can be obtained by the extremization of the action with respect to the
vector field $u^\mu$.  \par Let us define more precisely the
quantities appearing in the two extra Lagrangians in
Eq.~(\ref{lcor}). We first assume that the spacetime ${\cal M}$ is
globally hyperbolic so that it can be foliated as ${\cal
M}=\Sigma\times R$ where $\Sigma$ are three dimensional spacelike
hypersurfaces of constant $q$, where $q$ is a scalar. It follows that
the unit timelike vector field normal to these hypersurfaces is
$u_\mu\equiv-(\partial_\mu q)/(-g_{\alpha\beta} \partial^\alpha q
\partial^\beta q)^{1/2}$ which indeed satisfies $u_\mu u^\mu=-1$, with
respect to which we will define ``time'' and ``space''
components~\cite{carter,ellis}. Fr\"obenius theorem~\cite{Frob}
guarantees that the vector field $u_{\mu }$ is rotation-free and thus
that the field strength tensor can be written as $F_{\mu
\nu}=a_{\mu}\,u_{\nu}-a_\nu\,u_\mu$, $a_{\mu }$ being the acceleration
defined in Appendix~\ref{AppA}. As a consequence, if the vector field
$u_{\mu }$ is geodesic then $F_{\mu \nu}=0$. This will be for instance
the case in the FLRW case but does not need to be true in general. The
projector on the hypersurfaces $\Sigma$ defined by
\begin{equation}\label{def_perp}
\perp_{\mu\nu}\equiv g_{\mu\nu}+u_\mu u_\nu,
\end{equation}
coincides with the spatial metric as defined by an observer comoving
with $u_\mu$, since the line element can be rewritten as
\begin{equation}
\dd s^2=g_{\mu\nu}\dd x^\mu \dd x^\nu= -(u_\mu \dd x^\mu)^2
+\perp_{\mu\nu}\dd x^\mu \dd x^\nu.
\end{equation}
The covariant derivative $\DD_\alpha$ associated with the induced
3-metric of any tensor field $T_{\mu_1 \ldots \mu_n}^{\nu_1 \ldots
\nu_p}$, appearing in Eq.~(\ref{lcor}) defining ${\cal L}_{_{\rm cor}}$, is
defined as
\begin{equation}\label{8}
{\cal D}_\alpha T_{\mu_1 \ldots \mu_n}^{\nu_1 \ldots
\nu_p}\equiv\perp_{\mu_1}^{\lambda_1}\ldots
\perp_{\mu_n}^{\lambda_n}\perp^{\nu_1}_{\sigma_1}\ldots
\perp^{\nu_p}_{\sigma_p}\perp^\beta_\alpha\nabla_\beta T_{\lambda_1
\ldots \lambda_n}^{\sigma_1 \ldots \sigma_p},
\end{equation}
By construction, it is the covariant derivative associated with
$\perp_{\mu\nu}$ and is orthogonal to $u_\mu$, i.e.
\begin{equation}
{\cal D}_\alpha\perp_{\mu\nu}=0,\quad u^{\mu_i}{\cal D}_\alpha
T_{\mu_1\ldots \mu _i \dots \mu_n}^{\nu_1 \ldots \nu_p}
=u_{\nu_j}{\cal D}_\alpha
T_{\mu_1\ldots \mu_n}^{\nu_1 \ldots \nu _j\ldots \nu_p}
=u^\alpha{\cal D}_\alpha
T_{\mu_1\ldots \mu_n}^{\nu_1\ldots \nu_p}=0.
\end{equation}
Various identities satisfied by $u_\mu$, $\perp_{\mu\nu}$ and ${\cal
D}_\mu$, which are used repeatedly in the rest of this section are
given in Appendix~\ref{AppA}.  We finally define the operator ${\cal
D}^{2n}$ appearing in the Lagrangian Eq.~(\ref{lcor}) as
\begin{equation}
{\cal D}^{2n}\equiv{\cal D}_{\mu_1}{\cal D}^{\mu_1}\ldots{\cal
D}_{\mu_n}{\cal D}^{\mu_n}.
\end{equation}
For the particular case of a scalar field, this double derivative can
be written as
\begin{equation}\label{defd}
\DD_\mu\DD^\mu\phi=\perp_\mu^\alpha \nabla_\alpha\perp^\mu_\beta
\nabla^\beta\phi=\perp^{\alpha\beta}\nabla_\alpha\nabla_\beta\phi
+u^\alpha\nabla_\alpha\phi\nabla_\beta u^\beta.
\end{equation}
Throughout the paper, it will be understood that a derivative operator
applies directly and only on the first term appearing on its right;
derivatives of an ensemble of terms will be indicated using
brackets. We will also use the shorthand notation of an overdot for
the time derivative defined by an observer comoving with $u^\mu$,
i.e. $ \dot T_{\mu_1\ldots \mu_n}^{\nu_1\ldots \nu_p}\equiv
u^\alpha\nabla_\alpha T_{\mu_1\ldots \mu_n}^{\nu_1\ldots \nu_p}$. In
particular, $\dot\phi\equiv u^\alpha\nabla_\alpha\phi$. The above
double derivative operator can be shown to coincide with the
three-dimensional Laplacian as defined by an observer comoving with
$u^\mu$ [see Eq.~(\ref{A13})].  The corrective Lagrangian ${\cal
L}_{_{\rm cor }}$ thus contains only ``spatial'' derivatives. For
example, in Minkowski spacetime, this would yield plane wave solutions
to the field equations with a dispersion relation for the pulsation
$\omega^2$ as a series in powers of squared momentum $k^2$, where the
first term in $k^2$ results from the free Lagrangian and the higher
order terms come directly from the higher order Laplacians in ${\cal
L}_{_{\rm cor}}$.

\subsection{Stress--energy tensor}\label{sec_3}

The stress--energy tensor is obtained by varying the action
(\ref{action}) with respect to the metric
\begin{equation}
\delta S_{_\phi}=-\frac{1}{2}\int{\rm d}^4x\sqrt{-g}\, T_{\mu\nu}\delta
g^{\mu\nu} = \frac{1}{2}\int{\rm d}^4x\sqrt{-g}\, T^{\mu\nu}\delta
g_{\mu\nu}.
\end{equation}
In the following, we derive this stress-energy tensor and extract the
pressure and energy density under some hypotheses on the derivative
$\DD_\mu$. In Appendix~\ref{app_B}, we give a detailed calculation of
the stress-energy tensor in the simplest non-trivial case in which
only $b_{11}$ is non-vanishing, as considered by Jacobson and
Mattingly~\cite{jacobson,jacobson2}. The covariant derivative
obviously satisfies $\left[{\cal D}^2, g_{\mu\nu}\right]=0$ and we
further assume that it also satisfies
\begin{equation}\label{com2}
\left[{\cal D}^2, u^\mu\right]=
\left[{\cal D}^2, \nabla_\mu\right]=0.
\end{equation}
Although these requirements seem restrictive and may not apply in
general, they are fulfilled for the relevant cosmological case of the
FLRW spacetime we are interested in. We emphasize that these
requirements are not necessary to the derivation of the stress-energy
tensor in the case discussed in Appendix~\ref{app_B} in which only the
first coefficient $b_{11}$ is non-vanishing. The main point of the
calculation is the variation of the corrective term $S_{\rm
cor}$ to the action, which gives
\begin{eqnarray}
\delta S_{_{\rm cor}}&=&\frac{1}{2}\int {\cal L}_{_{\rm cor}}
g^{\mu\nu}\delta g_{\mu\nu}\sqrt{-g}\dd^4x
\nonumber
\\ 
&& - \sum_{n,p}b_{np}
\int \left\{\left[\sum_{i=0}^{n-1}\DD^{2i}(\delta\DD^2)
\DD^{2(n-i-1)}\phi\right]\DD^{2p}\phi +\DD^{2n}\phi
\left[\sum_{i=0}^{p-1}\DD^{2i}(\delta\DD^2)\DD^{2(p-i-1)}
\phi\right]\right\}\sqrt{-g}\dd^4x.
\end{eqnarray}
As a result of the commutation relations, the second integral can be 
rewritten through multiple integrations by parts as
\begin{equation}
\sum_{n,p} b_{np}(n+p)\int \left[\DD^{2(n+p-1)}\phi\right]
\delta\DD^2\phi\,\sqrt{-g}\dd^4x=\int E(\phi)\delta\DD^2\phi\,
\sqrt{-g}\dd^4x,
\end{equation}
where we used the short-hand notation $E(\phi)\equiv\sum
b_{np}(n+p){\cal D}^{2n+2p-2}\phi$. Using expressions
(\ref{d1}--\ref{d2}), we finally obtain the stress-energy tensor
\begin{eqnarray}\label{TT}
T_{\mu\nu} &=& \partial_\mu\phi\partial_\nu\phi
- \frac{1}{2}(\partial_\alpha\phi\partial^\alpha\phi)g_{\mu\nu}
-\sum _{n,p} b_{np}{\cal D}^{2n}\phi{\cal D}^{2p}\phi g_{\mu\nu}
+2\lambda u_\mu u_\nu \nonumber\\
& &+2\left[u_{(\mu}\nabla^\alpha u_{\nu)}\nabla_\alpha\phi
            -u_{(\mu}\nabla_{\nu)}u^\alpha\nabla_\alpha\phi
            -a_{(\mu}\nabla_{\nu)}\phi
            +{1\over 2}\perp_{\mu\nu}\Box\phi
         +{1\over 2}g_{\mu\nu}\left(\ddot{\phi}+\theta\dot{\phi}\right)
         \right]E(\phi)\nonumber\\
& & -2\nabla_{(\mu}E\nabla_{\nu)}\phi
    -2\dot{E}u_{(\mu}\nabla_{\nu)}\phi
    -2\dot{\phi}u_{(\mu}\nabla_{\nu)}E
    +\perp_{\mu\nu}\nabla_\alpha E\nabla^\alpha\phi
    +g_{\mu\nu}\dot E\dot\phi +T_{\mu \nu }^{\rm (F)},
\end{eqnarray}
where $T_{\mu \nu }^{\rm (F)}\equiv-4d_1F_{\mu \lambda }F_{\nu
}{}^{\lambda } +g_{\mu \nu}d_1F^{\alpha \beta }F_{\alpha \beta
}$. The expansion $\theta\equiv\perp^{\mu\nu}\nabla_\mu u_\nu$ is
defined in Appendix~\ref{AppA} together with the acceleration
$a^\mu$. To determine the Lagrange multiplier $\lambda$, we solve the
equation of motion for $u_\mu$ and obtain
\begin{equation}
\lambda = \frac{1}{2}\left[\left(\ddot\phi+\theta\dot\phi-2a^\alpha
          \nabla_\alpha\phi\right)E(\phi)-\dot\phi\dot{E}\right]
+\lambda ^{\rm (F)},
\end{equation}
where $\lambda ^{\rm (F)}\equiv2d_1u^{\mu }\nabla ^{\nu }F_{\nu \mu}$. Note 
that, as expected, $\lambda=0$ if the 
dispersion relation is
linear, which is obvious since in that case the action (\ref{action})
does not depend on the vector field $u^\mu$.
\par
To conclude this part, we derive the energy density and pressure as
defined by an observer comoving with $u_\mu$,
\begin{equation}
\rho \equiv u^\mu u^\nu T_{\mu\nu},\qquad p\equiv
\frac{1}{3}\perp^{\mu\nu}T_{\mu\nu},
\end{equation}
which gives
\begin{eqnarray}
\rho &=& \dot\phi^2+\frac{1}{2}\nabla_\alpha\phi\nabla^\alpha\phi
+\sum _{n,p}b_{np}{\cal D}^{2n}\phi{\cal D}^{2p}\phi 
+u^\mu u^\nu T_{\mu \nu }^{\rm (F)}+2\lambda ^{\rm (F)}, 
\\
3p &=& \dot\phi^2-\frac{1}{2}\nabla_\alpha\phi\nabla^\alpha
\phi
        - 3\sum _{n,p}b_{np} {\cal D}^{2n}\phi{\cal D}^{2p}\phi
        + \left(3{\cal D}^2\phi + a^\alpha\nabla\phi\right)E(\phi)
        +\nabla_\alpha E\nabla^\alpha\phi +\dot E\dot\phi -
\perp ^{\mu \nu}T_{\mu \nu }^{\rm (F)}.
\end{eqnarray}
Although the previous expressions hold for a non-interacting scalar
field, it is trivial to include a potential term in the above
equations. We now turn to the particular case of the FLRW metric.

\subsection{Friedmann-Lema\^{\i}tre-Robertson-Walker spacetime}
\label{sec_2c}

From now on, we restrict ourselves to a FLRW universe with flat
spatial sections, the metric of which is given by $g_{\mu\nu}\dd
x^\mu\dd x^\nu=-\dd t^2+a^2(t)\delta_{ij}\dd x^i\dd x^j$, where $a(t)$
is the scale factor and $t$ denotes the cosmic time. The Kr\"onecker 
symbol $\delta _{ij}$ represents the metric of constant time
hypersurfaces in cartesian coordinates. In this particular case 
the expansion of the universe provides a natural definition 
of space and time. We thus choose the
scalar function $q$ to be the cosmic time $t$ so that
$u_\mu=-\delta_\mu^0,\, u^\mu=\delta^\mu_0$. As a consequence, 
we now have $F_{\mu \nu}=0$. Also, the projector
$\perp_{\mu\nu}$ takes the simple form $\perp_{\mu\nu} \dd x^\mu\dd
x^\nu=a^2(t)\delta _{ij}\dd x^i\dd x^j$.
Using the previous equations, it can be checked that
\begin{equation}
\DD_\mu\DD^\mu\phi=\frac{1}{a^2}\delta^{ij}\partial_i\partial_j\phi\equiv
\frac{1}{a^2}\Delta\phi,
\end{equation}
i.e. $\DD_\mu\DD^\mu\phi$ is the four dimensional expression of the
three dimensional Laplacian, as expected from the general argument
(\ref{A13}). Let us note that the previous arguments can be easily
extended to a FLRW universe with non-flat spatial sections. In a FLRW
spacetime the commutation relations mentioned above are trivially
satisfied and the energy density and pressure reduce to
\begin{eqnarray}
\label{r1}
\rho &=& \frac{1}{2}\dot\phi^2+\frac{1}{2a^2}\delta^{ij}\partial_i\phi
\partial_j\phi + \sum _{n,p}b_{np}{\cal D}^{2n}\phi{\cal D}^{2p}\phi,
\\
\label{r2}
p &=& \frac{1}{2}\dot\phi^2-\frac{1}{6a^2}\delta^{ij}\partial_i\phi
\partial_j\phi+\sum _{n,p}b_{np}\left[{2\over 3}(n+p)-1\right] {\cal
D}^{2n}\phi{\cal D}^{2p}\phi ,
\end{eqnarray}
where we have implicitly integrated by parts and discarded a total
derivative term for the pressure. 

We now derive the field equation for $\phi$ to exhibit the modified
dispersion relation, and in particular to establish the link between
the series contained in the Lagrangian ${\cal L}_{_{\rm cor}}$ and the
Taylor expansion that defines the dispersion relation. From now on, we
use conformal time $\eta$ defined by ${\rm d}t\equiv a(t){\rm d}\eta$
and the reduced field $\mu \equiv a\phi$. Varying the action with
respect to $\mu$ gives
\begin{equation}
\mu''-\frac{a''}{a}\mu-\Delta\mu= -2a^2\sum _{n,p}
\frac{b_{np}}{a^{2(n+p)}}\Delta^{2(n+p)}\mu ,
\end{equation}
where a prime denotes a derivative with respect to $\eta$.  The last
term of this equation comes from the Lagrangian ${\cal L}_{_{\rm
cor}}$ which encodes all information on trans-Planckian physics. Since
this term is composed only of Laplacians, it gives rise to a series in
momentum, once we shift to Fourier space. For plane wave solutions of
the form $\mu(\eta ,{\bbox{x}}) = \mu_k(\eta)e^{i{\bbox{k}\cdot\bbox{x}}}$ one
obtains the equation
\begin{equation}\label{64}
\mu''_k+\left[\omega^2(k,\eta )-\frac{a''}{a}\right]\mu_k=0,
\end{equation}
with the modified dispersion relation
\begin{equation}
\omega^2(k)=k^2+2a^2\sum _{n,p}(-1)^{(n+p)}
b_{np}\left[\frac{k}{a(\eta)}\right]^{2(n+p)}.
\end{equation}
This dispersion relation can be rewritten in terms of physical
pulsation $\omega_{_{\rm phys}}=\omega/a$ and frequency $k_{_{\rm
phys}}=k/a$, as
\begin{equation}\label{disp_phys}
\omega^2_{_{\rm phys}}(k)=k_{_{\rm phys}}^2+2\sum _{n,p}(-1)^{(n+p)}b_{np}
k_{_{\rm phys}}^{2(n+p)}.
\end{equation}
Thus our procedure allows us to study the stress-energy tensor of a
quantum field with dispersion relation such that $\omega^2_{_{\rm
phys}}$ is an analytic function of $k^2_{_{\rm phys}}$.
\par
We now go one step further and second-quantize the system. We 
indeed study the behavior of gravitational
waves and density perturbations in the trans-Planckian regime, which
can be reduced in the pertubative approximation to the study of a free
minimally coupled quantum scalar field propagating in a FLRW classical
background, see e.g.~\cite{mukhanov92}.  We thus write the field
operator $\hat\mu$ (a hat denoting an operator) in the Heisenberg
representation as
\begin{equation}\label{63}
{\hat \mu}(\eta ,{\bbox{x}})=\int\frac{{\dd}{\bbox{k}}}{(2\pi)^{3/2}}
\left[\mu_k(\eta)\hbox{e}^{i{\bbox{k}\cdot\bbox{x}}}{\hat{\rm c}}_{\bbox{k}}+
\mu_k^*(\eta)\hbox{e}^{-i{\bbox{k}\cdot\bbox{x}}}{\hat{\rm c}}_{\bbox{k}}^\dag\right],
\end{equation}
where ${\hat{\rm c}}_{\bbox{k}}$ and ${\hat{\rm c}}_{\bbox{k}}^\dag$
represent annihilation and creation operators of quanta with momentum
${\bbox{k}}$ respectively, normalized as usual through
$\left[{\hat{\rm c}}_{\bbox{k}},{\hat{\rm
c}}_{\bbox{p}}^\dag\right]=\delta({\bbox{k}}-{\bbox{p}})$. The vacuum
state $\vert0\rangle$ is defined accordingly by ${\hat{\rm
c}}_{\bbox{k}}\vert0\rangle=0$ and a star denotes complex
conjugation. It is then easy to obtain the energy density and pressure
of the scalar field $\phi$ in its vacuum state by inserting
Eq.~(\ref{63}) into Eqs.~(\ref{r1}) and (\ref{r2}) to get
\begin{eqnarray}
\langle0|\hat\rho|0\rangle&=&\frac{1}{4\pi^2a^4}\int \dd k k^2
\left[a^2\left|\left(\frac{\mu_k}{a}\right)'\right|^2
+\omega^2(k)\left|\mu_k\right|^2\right],
\label{rhoeff}
\\
\langle0|\hat p|0\rangle&=&\frac{1}{4\pi^2a^4}\int \dd k k^2
\left[a^2\left|\left(\frac{\mu_k}{a}\right)'\right|^2 +\left(
\frac{2}{3}k^2\frac{\dd\omega^2}{\dd k^2}-\omega^2\right)
\left|\mu_k\right|^2\right].
\label{peff}
\end{eqnarray}
The above is one of the main results of the present article, and it
constitutes the basis of the discussion to follow in
Section~\ref{sec_4}. For the standard dispersion relation of massless
quanta $\omega(k)=k$, one recovers known results, namely that in the
limit $\mu_k\propto e^{-i\omega\eta}$, corresponding to subhorizon
modes (see Section~\ref{sec_4}), $p=\rho/3$ corresponding to a
relativistic fluid, while in the limit $(\mu_k/a)'\sim 0$,
corresponding to super-horizon modes (see Section~\ref{sec_4}), $p=
-\rho/3$.  It is of interest for what follows to note that even for
the standard dispersion relation, the fact that the quantity $\mu $ is
frozen on superhorizon scales does not imply that the corresponding
equation of state is of the cosmological constant type $p=-\rho$. 

For a modified dispersion relation, as mentioned before, the energy
density is the straightforward generalization of the standard
expression, with $\omega^2(k)$ denoting the modifed pulsation. However
the pressure expression is a non-trivial generalization, and the
presence of the term ${\rm d}\omega^2/{\rm d}k^2$ implies that various
equations of state may be obtained, depending on the shape of the
dispersion relation (this latter also determines the mode function
$\mu_k$). 

In the following, we will also be interested in the power spectrum of
the fluctuations of the field $\phi$, defined by
\begin{equation}
\label{defspec}
\langle 0\vert \phi ^2(\eta, {\bbox{x}}) \vert 0\rangle =\int _0^{+\infty }
\frac{{\rm d}k}{k}k^3P(k),
\end{equation}
where the power spectrum per logarithmic interval $k^3P(k)$ reads
\begin{equation}\label{41.2}
k^3P(k)=\frac{k^3}{2\pi ^2}\biggl \vert \frac{\mu _k}{a}\biggr\vert
^2.
\end{equation}
In the case of power law inflation and a standard dispersion relation,
the power spectrum is given by $k^3P(k)=A_{\rm S}k^{n_{\rm S}-1}$. In
the particular case of a de Sitter spacetime, the spectrum is scale
invariant, i.e. $n_{\rm S}=1$.

\section{Applications: equation of state and backreaction}\label{sec_4}

  In this section we apply our previous calculations of the
stress-energy tensor to two different situations that have arisen
recently in the literature. As a first application (\S~\ref{subsec_4.1})
we calculate the equation of state of gravitational waves of
super-Planck frequencies, which have been proposed to account for the
observed acceleration of the Universe~\cite{mbk}; our conclusions are
different and do not support the claim made by these authors. In
\S~\ref{subsec_4.2} and \S~\ref{subsec_4.3} we study the backreaction
problem of trans-Planckian modes in inflationary
cosmology~\cite{tanaka,starobinsky}. Our calculation is indeed well
suited to this case as it allows us to explicitly calculate the energy
density of the fluctuations and compare it to the background energy
density.

\subsection{Trans-Planckian and dark energy}\label{subsec_4.1}

It was proposed recently that gravitational waves of super-Planck
frequencies with a dispersion relation that exponentially decreases
beyond the cut-off (Planck) momentum would contribute significantly to
the present energy density, with an equation of state mimicking a
cosmological constant~\cite{mbk}. To discuss this claim, let us
consider the following dispersion relation
\begin{equation}\label{italiennes}
\omega^2_{_{\rm phys}}(k_{_{\rm phys}})=k_{_{\rm phys}}^2e^{-k_{_{\rm
phys}}^2/k_{\rm c}^2}\Longleftrightarrow
\omega^2(k)=k^2e^{-k^2/(a^2k_{\rm c}^2)},
\end{equation}
where $k_{\rm c}$ is the cut-off momentum. Even though it is not
exactly the dispersion relation considered in Ref.~\cite{mbk}, it is
simpler and very similar: it is linear at small momenta ($k\ll k_{\rm
c}$), reaches a maximum around $k_{\rm c}$ and decreases exponentially
at high momenta ($k\gg k_{\rm c}$), and it will serve our purposes
well enough (in addition, it was claimed that the result does not
depend on the details of the dispersion relation).

\begin{figure}[ht]
$$\epsfig{file=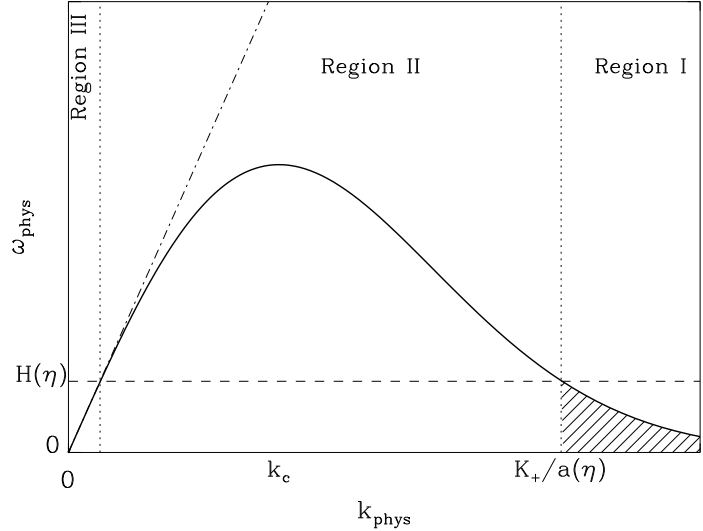, height=8cm}$$
\caption[]{Dispersion relation with an exponential decreasing tail, as
envisaged in Ref.~\cite{mbk}.  The dot-dashed line shows the linear
dispersion relation, and the thick solid curve shows the modified
dispersion relation. The dotted lines delimit three regions of
different evolutions for the mode function: in region I, corresponding
to the ``tail'', a mode has trans-Planckian momentum $k>ak_{\rm c}$
and sub-Hubble comoving pulsation $\omega(k)<aH$; in region II, a mode
has $\omega(k)>aH$, and in region III, modes are super-horizon sized
$k<aH$. The dashed area represents the ``tail'' modes defined by
$k>K_+$ (equivalently $k_{\rm phys}>K_+/a$ in the figure). }
\label{f0}
\end{figure}

In Fig.~\ref{f0} we show the qualitative behavior of the comoving
pulsation associated to this dispersion relation as a function of
comoving momentum and conformal time. There exists a region (``tail''
modes) in which the modes have $k_{\rm phys}>k_{\rm c}$ and a
pulsation smaller than the Hubble frequency $\omega_{\rm
phys}<H$. These modes are defined by $k>K_+>a k_{\rm c}$, where $K_+$ 
is obtained by equating $\omega$ with the comoving Hubble rate $a H$,
\begin{equation}\label{K+}
K_+\simeq ak_{\rm c}\sqrt{2\ln\frac{k_{\rm c}}{H}}.
\end{equation}
We further assume that the scale factor evolves as a power law of
conformal time, {\em i.e.}  $a(\eta)=a_0(\eta/\eta_0)^\beta$, where
$a_0$ is the dimensionless scale factor at time $\eta_0$, and
$\beta=1,2,-1$ respectively for radiation domination, matter
domination universe and de Sitter inflation. It is important to note
that a mode that is contained in the tail at time $\eta$ was already
contained in the tail at any previous time; indeed it can be easily
verified that the wavenumber $K_+$ increases with time for an
expanding scale factor and constant or decreasing curvature. In the
``tail'' region modes have a pulsation smaller than the Hubble
expansion rate, i.e. $\omega \ll aH=\beta/\eta$, which implies
$\omega^2 \ll a''/a = \beta(\beta-1)/\eta^2$. As a consequence, the
term $\omega^2$ in the equation of motion (\ref{64}) can be neglected
with respect to $a''/a$, so that the approximate solution to
Eq.~(\ref{64}) reads
\begin{equation}\label{sol1}
  \mu_k(\eta)\simeq C_+(k)a(\eta)+C_-(k)a(\eta)
  \int ^{\eta }\frac{\dd\eta'}{a^2(\eta')},
\end{equation}
where $C_+$ and $C_-$ are the coefficients of the growing and decaying
modes respectively. These two functions are inter-related through the
Wronskian normalization condition $\mu_k \mu_k^{*\prime} -
\mu_k^*\mu'_k = i$. The solution is indeed ``frozen'' since for the
growing mode the scalar field $\phi=\mu_k/a$ is constant in time. The
claim of Ref.~\cite{mbk} is based on the assumption that frozen modes
have an equation of state of a cosmological constant type
$p=-\rho$. As already mentioned above, this is not true even if the
dispersion relation is standard and the mode is frozen (super-horizon
sized), since the equation of state then reads $p=-\rho/3 $. In
addition, as we argue below, this claim does not hold either when the
dispersion relation is modified.

 In order to calculate the energy density and pressure, and the
relation between these two quantities, it is necessary to specify the
vacuum state. Unfortunately, for the ``tail'' modes, it cannot be
chosen unambiguously since the WKB approximation breaks down in the
limit $\omega\ll aH$ and the concept of adiabatic vacuum cannot be
used. This ambiguity also implies that the initial value of $\mu$,
i.e. the coefficients $C_-(k)$ and $C_+(k)$ cannot be determined
unambiguously. This problem has not been addressed in Ref.~\cite{mbk}
as only one branch of the solution to the equation of motion is
considered and the normalization coefficient (similar to the above
$C_\pm$) is taken to be independent of $k$.  In any case the solution
given in Ref.~\cite{mbk} is not a correct solution to the field
equation, since these authors neglect the contribution of the term
$a''/a$ in the limit $\eta\to-\infty$, but a direct comparison between
the various terms in their Eq.~(22) shows that instead the term
$a''/a$ dominates in this limit, in agreement with the above
discussion.

In the following, we propose to circumvent the ambiguity related to
the definition of a correct vacuum state by discussing two possible
generic sets of initial conditions.

\subsubsection{Power-law initial conditions}

The above problem is in a certain sense similar to the situation
encountered in cosmology before the advent of inflation. Indeed in the
absence of an inflationary epoch, quantum modes can only enter the
horizon as time goes, and the initial data had to be specified on
scales larger than the horizon in a regime where the field is
frozen. In contrast inflation has the virtue of stretching modes
beyond the horizon, so that the initial data can be specified without
ambiguity while the mode lies well inside the horizon. In the absence
of such a mechanism, it had been proposed to use a universal power law
to describe the power spectrum of the initial fluctuations, an
approach that relied on astrophysical arguments. We thus propose to
adopt a similar choice and take the growing mode coefficient to be a
power law in momentum
\begin{equation}
C_+(k)=k_{\rm I}^{-1/2}\left(\frac{k}{k_{\rm I}}\right)^\alpha,
\end{equation}
where $k_{\rm I}$ is a constant related to the amplitude of the
fluctuations and $\alpha$ a constant. It follows from
Eqs.~(\ref{rhoeff}--\ref{peff}) that the energy density and pressure
of the modes in the tail at a time $\eta$ are
\begin{eqnarray}
\langle0|\hat\rho|0\rangle &=& \frac{k_{\rm I}^{4}}{4\pi^2a^2}
\int_{K_+/k_{\rm I}}^\infty
u^{4+2\alpha}e^{-u^2\lambda^2} \dd u,
\\
\langle0|\hat p|0\rangle &=& -\frac{k_{\rm I}^{4}}{12\pi^2a^2}
\int_{K_+/k_{\rm I}}^\infty
u^{4+2\alpha}\left[1+2\lambda^2u^2\right]\,e^{-u^2\lambda^2} \dd u,
\end{eqnarray}
where we have introduced the dimensionless parameter $\lambda$ which
depends on the time at which the density and pressure are 
evaluated $\lambda\equiv k_{\rm I}/(ak_{\rm c})$. This can be 
integrated to give
\begin{eqnarray}
\langle0|\hat\rho|0\rangle &=& \frac{(ak_{\rm c})^4}{8\pi^2a^2}\left(
\frac{ak_{\rm c}}{k_{\rm I}}\right)^{2\alpha+1}\Gamma\left(\alpha+\frac{5}{2},
\frac{K_+^2}{a^2k_{\rm c}^2}\right),
\\
\langle0|\hat p|0\rangle &=& -\frac{(ak_{\rm c})^4}{8\pi^2a^2}\left(
\frac{ak_{\rm c}}{k_{\rm I}}\right)^{2\alpha+1}\left[\frac{1}{3}
\Gamma\left(\alpha+\frac{5}{2},\frac{K_+^2}{a^2k_{\rm c}^2}\right)
+\frac{2}{3}\Gamma\left(\alpha+\frac{7}{2},\frac{K_+^2}{a^2k_{\rm c}^2}\right)
\right],
\end{eqnarray}
where $\Gamma(\beta,x)\equiv \int _x^{\infty }{\rm d}t\,e^{-t}t^{\beta -1}$ is 
the incomplete gamma function \cite{GR}. The
equation of state parameter is thus obtained as (note that it does 
not depend on the arbitrary wave number $k_{\rm I}$)
\begin{equation}
\omega\equiv \frac{p}{\rho
}=-\frac{1}{3}-\frac{2}{3}\frac{\Gamma\left[\alpha+7/2,
K_+^2/(a^2k_{\rm c}^2)\right]}{\Gamma\left[\alpha+5/2,
K_+^2/(a^2k_{\rm c}^2)\right]},
\end{equation}
and since $K_+\gg ak_{\rm c}$, one can use the large argument expansion of the
incomplete gamma function $\Gamma(\beta,x)\simeq
x^{\beta-1}\exp{(-x)}$ at large $x$ to derive the following expression
\begin{equation}
\omega\simeq-\frac{1}{3}-\frac{2}{3}\alpha-\frac{4}{3}
 \ln\left(\frac{k_{\rm c}}{H}\right),
\end{equation}
where we have used Eq.~(\ref{K+}) to express $K_+/(ak_{\rm c})$. If $k_{\rm
c}$ is of order of the Planck scale, with $H_0/M_{\rm Pl}\simeq
10^{-61}$, then one obtains today
\begin{equation}
\omega\simeq -186.
\end{equation}
This value does not depend on the normalization scale $k_{\rm I}$ and
is in clear contradiction with a cosmological constant type equation
of state, {i.e.} $p\neq -\rho $. It is rather a ``phantom energy
component'' according to the terminology introduced in
Ref.~\cite{Cal}, {i.e.} a component for which $\omega <-1$. In
Ref.~\cite{Cal}, observational constraints on the equation of state
have been studied, and $\omega \simeq -186$ already appears in
contradiction with the SNIa data, see Fig.~6 of
Ref.~\cite{Cal}. Furthermore, if one evaluates the energy density
contained in the ``tail'' modes today (using again the asymptotic
value of the gamma function), one finds
\begin{equation}
\rho _{\rm tail} \sim H^2k_{\rm c}^2
\biggl(\frac{K_+}{k_{\rm I}}\biggr)^{2\alpha +1},
\end{equation}
or if we rather calculate $\Omega_{\rm tail}\equiv \rho_{\rm
tail}/\rho_{\rm crit}$, where $\rho_{\rm crit}\equiv3H^2M_{\rm
Pl}^2/8\pi$ is the critical energy density, one obtains
\begin{equation}
\Omega_{\rm tail} \simeq 0.1\left(\frac{k_{\rm c}}{M_{\rm
Pl}}\right)^2 \biggl(\frac{K_+}{k_{\rm I}}\biggr)^{2\alpha +1}.
\end{equation}
Therefore, if $k_{\rm c}\sim M_{\rm Pl}$, $\Omega_{\rm tail} \sim 1$
if $\alpha =-1/2$ and/or $k_{\rm I}=K_+\simeq 16.76 M_{\rm Pl}$. But
the important point is that this can be obtained only at the expense
of fine-tuning of the parameters, and the above choice does not seem
natural; actually, the parameter $k_{\rm I}$ is arbitrary and {\it a
priori} there exists no natural way to fix it.

\subsubsection{Minimising energy}

A second possibility is to fix the initial conditions by requiring
that the initial state configuration minimizes the energy, as
advocated in Ref.~\cite{BD} and used in Refs.~\cite{mb,bm}. To this
end we parametrize $\mu'_k$ as $\mu_k'/\mu_k\equiv x+iy$, where $x$
and $y$ are two real functions, and we rewrite the energy density as
\begin{equation}
\rho=\frac{1}{8\pi a^4}\int \dd kk^2\frac{1}{y}\left[{\cal
H}^2+x^2+y^2-2{\cal H}x+\omega^2\right]
\end{equation}
with ${\cal H}\equiv aH$.  One can now find the extrema of this
expression under the Wronskian normalization constraint to obtain the
mode function and its derivative at the initial time $\eta _{\rm i}$
\begin{equation}
|\mu_k(\eta_{\rm i})|=\frac{1}{\sqrt{2\omega(k,\eta_{\rm i})}},\qquad
\mu_k'(\eta_{\rm i})=[{\cal H}+i\omega(k,\eta_{\rm i})]\mu_k(\eta_{\rm i}).
\end{equation}
It follows that in the ``tail'' region, the field $\mu$ evolves in
time as
\begin{equation}
\mu_k^{[{\rm I}]}(\eta)=\frac{1}{\sqrt{2\omega(k,\eta_{\rm i})}}
\frac{a(\eta)}{a(\eta_{\rm i})}
\left\{1+i\omega(k,\eta_{\rm i})
\int_{\eta_{\rm i}}^\eta
\left[\frac{a(\eta)}{a(\eta_{\rm i})}\right]^2\dd\eta\right\}.
\end{equation}
The energy of the ``tail'' modes at a later time $\eta$ can then be
estimated as
\begin{equation}
\label{div}
\rho^{[{\rm I}]}(\eta) \simeq \frac{1}{8\pi^2a^2(\eta_{\rm
i})a^2(\eta)}\int_{K_+}^\infty \dd
kk^2\frac{\omega^2(k,\eta)}{\omega(k,\eta_{\rm i})}.
\end{equation}
It can then be checked that this quantity depends strongly on the
choice of the initial time. Moreover, since the dispersion relation
decreases exponentially fast as $k\to+\infty$ and $\eta_{\rm i}<\eta$,
the integral diverges exponentially when $k\rightarrow\infty$.

  In spite of the ambiguity related to the choice of the initial
state, the above two solutions have the merit to show that there is no
rigorous argument in favor of the claim made by Mersini {\it et
al.}~\cite{mbk}. In particular, one does not generically find (or
expect) the ``frozen'' modes of the ``tail'' neither to have a
cosmological constant type equation of state nor to have an energy
density today coinciding naturally with the critical energy
density. As we have mentioned above, the difference with the results
obtained in Ref.~\cite{mbk} presumably lies in an incorrect assumption
made by these authors to derive the time evolution of the mode
functions. One should also add that these authors~\cite{mbk} have not
addressed the problem of the energy contained in modes with momentum
$k<K_+$ (in particular region II in Fig.~\ref{f0}). Indeed if one
considers the ``tail'' modes as a source of gravitational energy
today, there is {\it a priori} no reason to discard the contribution
of other modes. However the subhorizon non-frozen modes oscillate, and
it is easy to see that their energy density will be of order $M_{\rm
Pl}^4$, which is nothing less that the celebrated long-standing
problem of the cosmological constant.

\subsection{Backreaction: general discussion}\label{subsec_4.2}

The previous section has briefly touched upon the issue of
backreaction in trans-Planckian cosmology, which arises whenever the
energy density of the fluctuations (scalar or tensor) becomes
comparable to the background energy density. In this case, the
perturbative approximation used to describe the evolution of the mode
functions breaks down, and one has to resort to a second order
(complex) treatment of the Einstein equations~\cite{ABM}.  \par
Backreaction effects may arise in different situations in inflationary
cosmology. Liddle and Lyth~\cite{lyli} (in collaboration with
E.~Stewart) pointed out that if the modes of comoving wavelength of
order of the horizon size, i.e. modes of cosmological interest for
structure formation, start in a non-vacuum initial state, then
inflation suffers from a very important backreaction problem. It is
interesting for the following discussion to summarize briefly their
argument. A mode with momentum $k$ exits the horizon $N_{\rm e}(k)\sim
56 - \ln(k/H_0)$ $e$-folds before the end of inflation~\cite{KT91}.
At horizon crossing, this mode carries a physical momentum $k_{\rm
phys}=H_{\rm inf}$, $H_{\rm inf}\sim10^{13}\,$GeV being the Hubble
scale during inflation.  Therefore, this mode had a physical momentum
$k_{\rm phys}= \exp(N_{\rm i})H_{\rm inf}$ at the onset of inflation,
$N_{\rm i}=N_{\rm tot}-N_{\rm e}$ being the number of $e$-folds of
inflation before horizon crossing, and $N_{\rm tot}$ the total number
of $e$-folds of inflation. There is no reason to expect that inflation
lasted just sufficiently long so as to stretch beyond the horizon only
those modes with wavelength smaller than the horizon size today, and
in general $N_{\rm tot}\gg N_{\rm e}$ hence $N_{\rm i} \gg 1$. The
energy density contained in these modes at the begining of inflation
is thus $\delta\rho_k \sim (2\pi^2)^{-1}n_k\exp(4N_{\rm i})H_{\rm
inf}^4$ up to a numerical factor of order unity, with $n_k$ the
occupation number (we discard the zero-point energy). The ratio of
this energy density to the critical energy density thus reads
$\delta\rho/\rho_{\rm crit} \sim n_k\exp(4N_{\rm i}) (H_{\rm
inf}/M_{\rm Pl})^2$; finally $H_{\rm inf}\sim 10^{-6} M_{\rm Pl}$
implies the very stringent constraint $n_k\lesssim \exp(-4N_{\rm i} +
28)$ to avoid backreaction problems. In other words, either $N_{\rm
i}\lesssim 7$, which may seem unnatural, or $n_k\ll1$ which implies
that these modes are in their vacuum state or very close to it. Even
though this argument was developed in the context of standard
inflationary cosmology without modified dispersion relation and
trans-Planckian physics, it can be carried over directly to our
present problem.

In fact, this has been done recently by Tanaka~\cite{tanaka} who
assumed that the modification of the dispersion relation could be
represented by a non-vacuum initial state in a theory with unmodified
dispersion relation. This implied, in much the same way as above, that
backreaction effects should be very
important. Starobinsky~\cite{starobinsky} has further argued that if
adiabaticity is broken at some point in the evolution of a given mode,
then at late times, when adiabaticity is restored, this mode behaves
as the sum of positive and negative energy plane waves, meaning that a
finite amount of quanta has been created, and the energy density they
carry is large compared to the critical energy density today. However
neither of these authors had at their disposal a rigorous expression
for the energy density and they could not provide a quantitative
estimate of this backreaction energy density. Nevertheless as we show
in the following their conclusions remain correct, notably because the
correct expression for the energy density Eq.~(\ref{rhoeff}) is the
generalization of the standard expression. The overall picture for the
origin and magnitude of backreaction can be described as follows.

  Consider a general dispersion relation such that for a given
comoving wavenumber $k$, for time $\eta_2(k)<\eta<\eta_3(k)$, where
$\eta_3(k)\simeq-1/k$ denotes the conformal time of horizon crossing,
the WKB approximation is valid, i.e. $\omega'/\omega^2\ll1$. This
notably implies $\omega(k)\gg {\cal H}$ for
$\eta_2(k)<\eta<\eta_3(k)$, as the WKB condition would not be
satisfied otherwise. In this case, the mode function evolves according
to
\begin{equation}
\label{solgenewkb}
\mu_k(\eta)\simeq \frac{\alpha(k)}{\sqrt{2\omega(k,\eta)}}e^{-i\int
^{\eta }\omega (k,\eta '){\rm d}\eta '} +
 \frac{\beta(k)}{\sqrt{2\omega(k,\eta)}}e^{i\int ^{\eta }
\omega (k,\eta '){\rm d}\eta '},
\end{equation}
with $\vert\alpha(k)\vert^2 - \vert\beta(k)\vert^2=1$ from the
Wronskian normalization condition. In principle the Bogoliubov
coefficients $\alpha(k)$ and $\beta(k)$ depend on time, but their time
dependence can be neglected to first order in the WKB expansion.
These coefficients can be obtained by matching the mode function
$\mu_k$ and its first derivative at time $\eta_2(k)$ with the solution
of the field equation in the region $\eta < \eta_2(k)$. If the WKB
approximation was also valid at all times $\eta < \eta_2(k)$, and the
Bunch-Davies adiabatic vacuum is chosen as the initial state of the
field (in mode $k$ at least), then $\beta_k\approx0$ and
$\alpha_k\approx1$ at time $\eta$. Furthermore, the modification to
the power spectrum in this case is of first order in $\beta_k$, and
thus remains small~\cite{np}.  However, if at some time prior to
$\eta_2(k)$, the WKB condition was violated, then $\vert\beta_k\vert$
can be {\it a priori} large; this had been remarked by
Starobinsky~\cite{starobinsky}. This may induce large modifications in
the power spectrum, but it also represents the creation of a large
amount of energy density due to the breaking of adiabaticity. In
effect, the energy density contained at time $\eta$ in all modes such
that the mode function $\mu_k(\eta) $ is given by Eq.~(\ref{solgenewkb}) 
can be written as
\begin{eqnarray}
\langle 0 \vert\hat\rho\vert 0 \rangle (\eta ) &=&
\frac{1}{4\pi^2a^4}\int {\rm d}k k^2\biggl\{
\frac{1}{2\omega }
\biggl[\omega ^2 +\vert \gamma \vert ^2\biggr]+
\frac{\vert \beta _k\vert ^2}{\omega }
\biggl[\omega ^2 +\vert \gamma \vert ^2\biggr]
+\frac{\alpha _k\beta _k^*}{2\omega }\biggl[\omega ^2 +\gamma ^2\biggr]
e^{-2i\int ^{\eta }\omega (k,\eta '){\rm d}\eta '}
\nonumber \\
& & \qquad\quad
+\frac{\alpha _k^*\beta _k}{2\omega }\biggl[\omega ^2 +(\gamma ^*)^2\biggr]
e^{2i\int ^{\eta }\omega (k,\eta '){\rm d}\eta '}\biggr\},
\end{eqnarray}
where we have used $\vert \alpha _k\vert ^2=1+\vert \beta _k\vert
^2$. Again, note that the above integral only covers the range of
comoving momenta for which Eq.~\ref{solgenewkb}) is valid. In
principle, the total energy density should also contain the
contribution of other domains of momenta. In the above expression, the
quantity $\gamma $ is defined as follows
\begin{equation}\label{defg}
\gamma (k,\eta )\equiv\left[\frac{\omega'(k,\eta)}{2\omega(k,\eta)}
+i\omega(k,\eta)+\frac{a'}{a}\right].
\end{equation}
In a situation where WKB is a good approximation, we have $\gamma
/\omega \simeq i$ and the previous expression reduces to
\begin{equation}\label{rho2}
\langle0\vert\hat\rho\vert0\rangle=\frac{1}{4\pi^2a^4}\int \dd k k^2
\left(\frac{1}{2} + \vert\beta_k\vert^2\right)\omega(k).
\end{equation}
 Note that in order to remove the two oscillatory terms, no procedure
of time averaging is needed in contrast with what has been done in
Ref.~\cite{tanaka}. If $\omega(k)$ does not decrease faster than
$k^{-3}$ as $k\to+\infty$, this integral diverges as in flat space due
to the first term in the integrand. We interpret this infinite
quantity as the zero-point energy which we subtract (adiabatic
regularization), and interpret the remainder as the presence of finite
energy density in modes with occupation number $\vert\beta_k\vert^2$,
as usual. It is interesting to note that if $\omega(k)$ decreases
faster than $k^{-3}$ as $k\to+\infty$, the integral is no longer
divergent, and delicate questions on the necessity of renormalization
arise. We will not touch upon these subtle issues in the present
article, and will adopt the simplistic point of view in which the
zero-point energy is subtracted in all cases. In Ref.~\cite{tanaka},
an expression similar to Eq.~(\ref{rho2}) above had been used to
discuss the magnitude of backreaction, but the origin of $\beta_k$ had
been left unspecified. The above discussion establishes the link that
was missing and it also justifies more rigorously the approach of
Refs.~\cite{tanaka,starobinsky}, {i.e.}  it shows explicitely that
having a non standard dispersion relation is equivalent to considering
non-vacuum quantum states for the perturbations as guessed in these
studies. Furthermore, it also shows how to calculate $\beta_k$, {i.e.}
through the matching with the solution to the mode equation at time
$\eta<\eta_2(k)$.  \par In the following, we provide a concrete
example of such a calculation and discuss the magnitude of $\beta_k$
for a general class of dispersion relation. The WKB approximation is
valid whenever $\vert\omega'/\omega^2\vert\ll 1$, or
\begin{equation}
\left\vert\frac{H}{\omega_{\rm phys}}\left(1-\frac{{\rm d}\ln
\omega_{\rm phys}}{{\rm d}\ln k_{\rm
phys}}\right)\right\vert\ll1.
\end{equation}
The WKB approximation can thus be violated in two general ways: by
space-time curvature effects, {i.e.} when $\omega_{\rm phys}\ll
H$, and/or by singularities in the dispersion relation or its first
derivative. In the following we will be interested only in the former
class. An illustrative example of the latter is given by the
Corley-Jacobson dispersion relation that becomes complex, which has
already been discussed extensively~\cite{mb,bm}. We note that the
dispersion relation considered in Section~\ref{subsec_4.1} violates
the WKB approximation at large physical momenta in both ways, namely
${\rm d}\ln\omega_{\rm phys}/{\rm d}\ln k_{\rm phys}$ diverges as 
$k_{\rm phys}\to+\infty$ (or $\eta\to-\infty$), and $\omega_{\rm
phys}$ decreases much faster than $H$ as $\eta\to-\infty$. 
In order to compute the Bogoliubov coefficient $\beta_k$ one needs to
solve the mode equation in the region where the WKB approximation is
violated. However, as discussed in Section~\ref{subsec_4.1}, in this
region one cannot determine unambiguously a vacuum state, and in
particular one cannot fix unambiguously initial data. We thus make a
further assumption, and assume that the dispersion relation is such
that the WKB approximation is restored in the far past
$\eta\to-\infty$. More precisely we assume that for a given comoving
wavenumber $k$, there exists a time $\eta_1(k)$ such that for
$\eta<\eta_1(k)$, the WKB condition is satisfied; this allows us to
define proper initial data for the mode evolution. 

\subsection{Backreaction: examples}\label{subsec_4.3}

\begin{figure}[ht]
\centerline{\epsfig{file=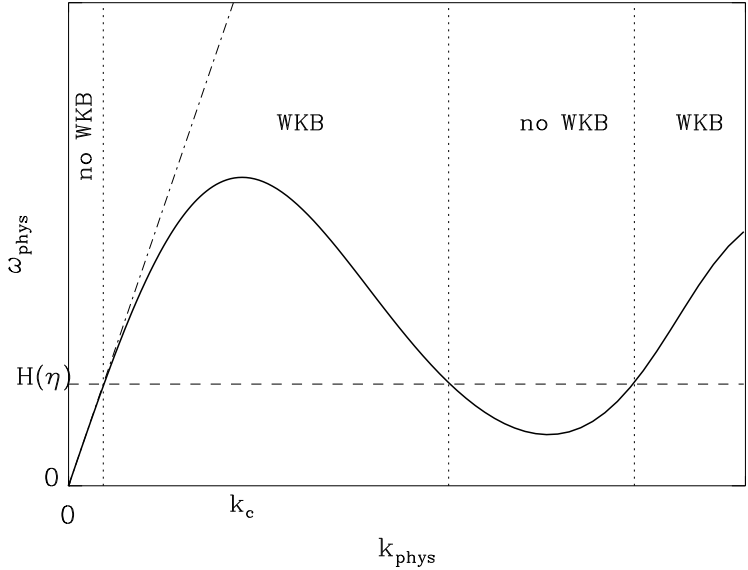, width=12cm}}
\caption{Example of a modified dispersion relation which breaks the
WKB approximation in the trans-Planckian regime when $\omega_{\rm
phys}<H$, as indicated.}
\label{fig1}
\end{figure}

A dispersion relation with the above property can be constructed
easily. For instance, consider the lowest order generalization of the 
Corley-Jacobson dispersion relation, {i.e.}  $\omega ^2_{_{\rm
phy}} =k^2_{_{\rm phys}}+2b_{11}k^4_{_{\rm phys}}-2b_{12}k^6_{_{\rm
phys}}$. Depending on the sign of the coefficients $b_{11}$ and
$b_{12}$, this dispersion relation can present both a maximum around
$k_{\rm c}$ and a minimum at physical momentum larger than $k_{\rm
c}$. For a specific choice of these coefficients, this minimum can be
chosen to be smaller than the Hubble scale at some time $\eta$, as
depicted in Fig.~\ref{fig1}.

This is only an example of the desired class of dispersion relations,
and the following discussion remains general with respect to the form
of $\omega_{\rm phys}(k_{\rm phys})$. We fix the initial conditions 
in the region $\eta<\eta_1(k)$ where
the WKB approximation holds by selecting positive frequency modes,
which corresponds to the choice of the Bunch-Davies adiabatic vacuum
state for the field,
\begin{equation}
\mu_k^{[0]}(\eta)=\frac{1}{\sqrt{2\omega(k,\eta)}}\hbox{e}^{-i
\int_{\eta_{\rm i}}^\eta\omega(k,\eta')\dd\eta'}.
\end{equation}
At time $\eta_1(k)$, this mode enters region I where
$\omega(k)\ll{\cal H}$, and $\mu_k$ and its first derivative must be
matched to the solution Eq.~(\ref{sol1}), which gives
\begin{equation}
\mu_k^{[{\rm
I}]}(\eta)=\frac{1}{\sqrt{2\omega[k,\eta_1(k)]}}\hbox{e}^{-i
\int_{\eta_{\rm
i}}^{\eta_1(k)}\omega(k,\eta')\dd\eta'}\frac{a(\eta)}{a(\eta_1)}
\left\lbrace1-\gamma (k,\eta
_1)\int_{\eta_1(k)}^\eta\left[\frac{a(\eta_1)}{a(\eta')}
\right]^2\dd\eta'\right\rbrace ,
\end{equation}
where the quantity $\gamma$ has been defined previously, see
Eq.~(\ref{defg}). Note that $\gamma_1$ carries dimensions of inverse
time and in particular, for general ${\cal C}^1$ dispersion relations,
$\gamma_1\eta_1$ is a number of order unity, which depends weakly on
$k$ [see Eq.~(\ref{gameta}) below]. At time $\eta_2(k)$, the mode
enters region II in which the WKB approximation holds,
i.e. $\omega(k)\gg{\cal H}$, and $\mu_k$ and its first derivative must
be matched to the WKB solution, giving
\begin{equation}
\mu_k^{[{\rm II}]}(\eta)=\frac{1}{\sqrt{2\omega(k,\eta)}}
\left[\alpha_k\hbox{e}^{-i\int_{\eta_2(k)}^{\eta}\omega(k,\eta')\dd\eta'}
+\beta_k\hbox{e}^{+i\int_{\eta_2(k)}^{\eta}\omega(k,\eta')\dd\eta'}
\right],
\end{equation}
and the two coefficients $\alpha_k$ and $\beta_k$ are explicitely
given by
\begin{eqnarray}
\label{expal}
\alpha_k&=&\frac{i}{2}\frac{\hbox{e}^{-i\int_{\eta_i}^{\eta_1(k)}
\omega(k,\eta') \dd\eta'}}
{\sqrt{\omega[k,\eta_1(k)]\omega[k,\eta_2(k)]}}
\left[\frac{a_2}{a_1}\gamma_2^*-\frac{a_1}{a_2}\gamma_1-
a_1a_2\gamma_1\gamma_2^*\int_{\eta_1(k)}^{\eta_2(k)}
\frac{\dd\eta'}{a^2}\right] ,\\
\label{expbeta}
\beta_k&=&-\frac{i}{2}\frac{\hbox{e}^{-i\int_{\eta_i}^{\eta_1(k)}
\omega(k,\eta') \dd\eta'}}
{\sqrt{\omega[k,\eta_1(k)]\omega[k,\eta_2(k)]}}
\left[\frac{a_2}{a_1}\gamma_2-\frac{a_1}{a_2}\gamma_1-
a_1a_2\gamma_1\gamma_2\int_{\eta_1(k)}^{\eta_2(k)}
\frac{\dd\eta'}{a^2}\right],
\end{eqnarray}
and where $\gamma_2$ is defined as in Eq.~(\ref{defg}) albeit it is
evaluated at time $\eta_2(k)$. The previous equations give the general
expressions of the coefficients $\alpha _k$ and $\beta _k$ in region
II. These expressions can be further worked out in the following two
situations: (i) when $|\eta_1|\gg|\eta_2|$ and the scale factor is
written as $a(\eta) \equiv a_{\rm inf}(\eta/\eta_{\rm inf})^\beta$ or
(ii) when $\eta_2\sim\eta_1$.  \par Let us start with the first
situation. Assuming $|\eta_1|\gg|\eta_2|$ and using
$\omega_2\eta_2=\omega_1\eta_1=-\sqrt{\beta(\beta-1)}$, the leading
contribution to the two coefficients $\alpha_k$ and $\beta_k$ can be
written as
\begin{eqnarray}
\alpha_k&\simeq&-\frac{i}{2} \frac{e^{-i\int_{\eta_{\rm
      i}}^{\eta_1(k)}\omega(k,\eta')
      \dd\eta'}}{\sqrt{\beta(\beta-1)}}\gamma_2^*\vert \eta_2\vert
      \left(1+\frac{\gamma_1\eta_1}{1-2\beta}\right)
      \left(\frac{\eta_2}{\eta_1}\right)^{\beta-1/2} ,\\
\beta_k&\simeq&\frac{i}{2} \frac{e^{-i\int_{\eta_{\rm
      i}}^{\eta_1(k)}\omega(k,\eta')
      \dd\eta'}}{\sqrt{\beta(\beta-1)}}\gamma_2\vert \eta_2\vert
      \left(1+\frac{\gamma_1\eta_1}{1-2\beta}\right)
      \left(\frac{\eta_2}{\eta_1}\right)^{\beta-1/2},\label{leading}
\end{eqnarray}
The values of $\gamma_1\eta_1$ and $\gamma_2\eta_2$ can be cast in the
form
\begin{equation}
\label{gameta}
(\gamma \eta )_{1,2}=\frac{3}{2}\beta - i\sqrt{\beta (\beta -1)}
-\frac{1}{2}\beta \frac{{\rm d}\ln \omega _{_{\rm phys}}}
{{\rm d}\ln k_{_{\rm phys}}}\biggl \vert _{1,2},
\end{equation}
which shows that these quantities are of order unity and do not depend
strongly on $k$ unless the dispersion relation contains exponential
factors important at the time of matching. It follows that in region
II, one has
\begin{equation}
\left|\mu_k^{[{\rm II}]}(\eta )\right|^2\simeq
\frac{1}{4\omega(k,\eta)\beta(\beta-1)}
\left(\frac{\eta_2}{\eta_1}\right)^{2\beta-1}\left|\gamma_2\eta_2\right|^2
\left|1+\frac{\gamma_1\eta_1}{1-2\beta}\right|^2\biggl\{1-
\cos\biggl[2\varphi_2+
2\int_{\eta_2(k)}^\eta\omega(k,\eta')\dd\eta'\biggr]\biggr\},
\end{equation}
where $\varphi _2\equiv \arg (\gamma _2)$. In region III, in which the
mode exits the horizon on the linear part of the dispersion relation,
the solution reads $\mu_k(\eta )\simeq C_{\rm III}(k)a(\eta )$. The
constant $C_{\rm III}(k)$ is obtained by matching the super-horizon
solution with the solution in region II given above and it essentially
determines the power spectrum [see Eq.~(\ref{41.2})]. When the mode
$k$ exits the zone II and enters the zone III, the dispersion relation
is linear $\omega\simeq k$. We deduce that
$\eta_3(k)=-\sqrt{\beta(\beta-1)}/k$ and thus that
\begin{eqnarray}
\label{psfinal}
k^3P(k) & = &\frac{\eta_{\rm inf}^{2\beta }}{8\pi ^2a_{\rm
inf}^2}[\beta (\beta -1)]^{-\beta -1} \left|\gamma_2\eta_2\right|^2
\left|1+\frac{\gamma_1\eta_1}{1-2\beta}\right|^2 k^{2\beta
+2}\biggl(\frac{\eta_2}{\eta_1}\biggr )^{2\beta-1} \nonumber \\ & &
\times \biggl\{1- \cos\biggl[2\varphi_2+
2\int_{\eta_2(k)}^{-\sqrt{\beta(\beta-1)}/k}\omega(k,\eta')
\dd\eta'\biggr]\biggr\}.
\end{eqnarray}
Two interesting features should be noted on this form of the power
spectrum. First, the standard spectral index of the overall amplitude
$n_{\rm S}-1=2\beta +2$ is modified due to the factor
$(\eta_2/\eta_1)^{2\beta-1}$ which {\it a priori} depends on
$k$. Second, the power spectrum exhibits superimposed oscillations and
therefore it can {\it a priori} vanish for some values of $k$. The
above formula can be used to study the features of these oscillations
once the dispersion relation has been specified but we will not pursue
this goal here.  However, in the particular case of a de Sitter
inflationary period $(\beta=-1$), the previous features are no longer
present since all quantities are functions of $k\eta$ only. Notably,
since $H$ is a constant in de Sitter space, the two solutions $\eta_1$
and $\eta_2$ to the equation $\omega_{\rm phys}=H$ read $\eta_1=
C_1/k$ and $\eta_2=C_2/k$, where $C_1$ and $C_2$ are constants, and
therefore $\eta_2/\eta_1$ is independent of $k$. For the same reason,
since $\omega(k,\eta)=\omega(k\eta)$, a change of variables $\eta\to
u\equiv k\eta$ in the integral appearing in Eq.~(\ref{psfinal}) shows
that this quantity does not depend on $k$. The spectral index is thus
unchanged, i.e. $n_{\rm S}=1$, and the superimposed oscillations
reduce to a constant numerical factor. The power spectrum is thus
unchanged except for an overall modification of its amplitude; this
latter is actually magnified by a factor ${\cal
O}\left[(\eta_1/\eta_2)^3\right]$ ($\beta=-1$ for de Sitter). This
result is in agreement with Refs.~\cite{KN01,EGK01} where the power
spectrum was calculated for a theory with modified canonical
commutation relations but unmodified dispersion relation. In
Ref.~\cite{KN01}, it was argued that the power spectrum spectral index
remains unchanged for de Sitter inflation, and in Ref.~\cite{EGK01} it
was further argued that only the overall amplitude of the power
spectrum is affected.
Finally, the energy density contained in the modes in region II at
time $\eta$ can be written as:
\begin{equation}
\rho (\eta ) \simeq\frac{1}{16\pi^2\beta (\beta -1)a^4}\int_{\cal K} {\rm d}k\,k^2
\left(\frac{\eta_2}{\eta_1}\right)^{2\beta-1}\left|\gamma_2\eta_2\right|^2
\left|1+\frac{\gamma_1\eta_1}{1-2\beta}\right|^2
\omega(k,\eta),
\end{equation}
and the domain of integration ${\cal K}$ is such that $aH < k < a k_2$
with $k_2$ being the smallest wavenumber such that $\omega_{\rm
phys}(k_2)=H$ and $k_2>H$. It is artificial to consider different
domains of wavenumbers and to treat the corresponding contributions to
the energy density separately, but those modes with wavenumbers $k > a
k_1$ [where $k_1$ is the other wavenumber $>H$ such that $\omega_{\rm
phys}(k_1)=H$] are in their vacuum state and WKB holds so the
contribution to the integral has been removed, while those with $a k_1
< k < ak_2$ are in the non-WKB zone, where it is ambiguous to define a
vacuum state and to calculate the contribution to the energy
density. On the other hand, it is sufficient to show that one of these
contributions is of the order of the background energy density to
demonstrate that there is a backreaction problem. This is the spirit
of the following calculation and the interval chosen is particulary
well-suited since in the standard case the corresponding contribution
vanishes. In the present case, one obtains for the case $n_{\rm S}=1$
\begin{equation}
\rho (\eta )\sim \biggl(\frac{\eta _1}{\eta _2}\biggr)^3
\int_{\cal K} {\rm d}k_{\rm phys}\,k^2_{\rm phys}
\omega _{\rm phys }(k)=\biggl(\frac{\eta _1}{\eta _2}\biggr)^3
{\cal O}(k_{\rm c}^4),
\end{equation}
where we have approximated the integral to the peak value of 
$\omega_{\rm phys}$ at $k_{\rm c}$. This value should be 
compared with the background energy density during 
inflation $\rho _{\rm inf}=M_{\rm Pl}^2H_{\rm inf}^2$. Therefore, 
if we take $k_{\rm c}\simeq M_{\rm Pl}$, we see that $\rho \gg 
\rho _{\rm inf}$ due to $H_{\rm inf}<M_{\rm Pl}$ and 
$\eta _1/\eta _2 \gg 1$, the latter equation expressing the fact 
that the coefficient $\beta _k$ is large in region II. This 
result coincides with the analysis of 
Refs.~\cite{tanaka,starobinsky}. 
\par
The other situation where the spectrum and the energy density 
can be evaluated explicitely if the time $\eta _2$ is 
close to $\eta _1$. If we write $\eta _2=\eta _1(1+\epsilon )$, 
where $\epsilon $ is a small parameter, then one finds for the 
two coefficients $\alpha _k$ and $\beta _k$ 
[see Eqs.~(\ref{expal}, \ref{expbeta})] 
\begin{eqnarray}
\alpha _k &\simeq & \biggl\{1-\epsilon \frac{i\omega _1\eta _1}{2}
\biggr[1+\frac{Q_1}{\omega ^2_1}-\frac{\kappa a^2_1}{2\omega ^2_1}
\biggl(\frac{\rho _0}{3}-p_0\biggl)\biggr]\
\biggr\}
e^{-i\int_{\eta_{\rm i}}^{\eta_1(k)}\omega(k,\eta')
\dd\eta'} +{\cal O}(\epsilon ^2), 
\\
\beta _k &\simeq & -\epsilon \frac{i\omega _1\eta _1}{2}
\biggr[1-\frac{Q_1}{\omega ^2_1}+\frac{\kappa a^2_1}{2\omega ^2_1}
\biggl(\frac{\rho _0}{3}-p_0\biggl)\biggr]
e^{-i\int_{\eta_{\rm i}}^{\eta_1(k)}\omega(k,\eta')
\dd\eta'}+{\cal O}(\epsilon ^2),
\end{eqnarray}
where $\rho _0$ and $p_0$ are the background energy density and
pressure at time $\eta_1(k)$ respectively. The coefficient $Q$ is
defined by $Q\equiv 3\omega '^2/(4\omega ^2)-\omega ''/(2\omega
)$. The interpretation of these formulae is as follows. To leading
order in $\epsilon $, the coefficient $\alpha _k$ is just a pure phase
and the coefficient $\beta _k$ vanishes as expected. Two terms appear
to the next order in $\epsilon $ where the corrections show up. The
first is $Q/\omega ^2$. The condition for WKB to be valid is $ \vert
Q/\omega ^2 \vert \ll 1$ and therefore this term indicates the
magnitude of violation of the WKB approximation. The second term is
proportional to $\rho _0/3-p_0$.  The presence of this term is also
natural because it vanishes for radiation ($a''=0$), in which case the
exact solution for the mode function $\mu _k(\eta )$ is a complex
exponential and no correction is expected for the overall
amplitude. Finally, repeating the same calculations as above for the
power spectrum, one finds
\begin{eqnarray}
\label{psfinalepsi}
k^3P(k) & = & \frac{\eta_{\rm inf}^{2\beta }}{4\pi ^2a_{\rm inf}^2}[\beta
(\beta -1)]^{-\beta }k^{2\beta +2} \nonumber \\
& & \times \biggl\{1 -\epsilon \omega
_1\eta _1 \biggr[1-\frac{Q_1}{\omega ^2_1}+\frac{\kappa
a^2_1}{2\omega ^2_1} \biggl(\frac{\rho _0}{3}-p_0\biggl)\biggr] \sin
\biggl[ 2\int_{\eta_2(k)}^{-\sqrt{\beta (\beta -1)/k}}\omega(k,\eta')
\dd\eta'\biggr]\biggr\} +{\cal O}(\epsilon ^2).
\end{eqnarray}
The correction to the power spectrum is of order $\epsilon $ as
expected. To this order, as before, we have a modified overall
amplitude and superimposed oscillations appear. We can also estimate
the energy density. Inserting the expression of the coefficient $\beta
_k$ into Eq.~(\ref{rho2}), one obtains
\begin{equation}
\rho \sim {\cal O}(\epsilon ^2k_{\rm c}^4).
\end{equation}
Therefore, there is no backreaction problem if $\epsilon ^2k_{\rm
c}^4\lesssim M_{\rm Pl}^2H_{\rm inf}^2$. Let us write $k_{\rm c}$ as
$k_{\rm c}\equiv 10^{-s}M_{\rm Pl}$, where the coefficient $s$ fixes
the scale of the characteristic wavenumber with respect to the Planck
mass. One can also write $\epsilon $ as $\epsilon \equiv 10^{-p}$,
where $p$ roughly gives the order of magnitude of the modification to
the power spectrum. If we take $H_{\rm inf}=10^{-6}M_{\rm Pl}$, then
there is no backreaction problem if $p+2s\gtrsim 6$ that is to say if
$s\gtrsim 3-p/2$. Therefore a modification of the spectrum of order
$10\%$ in principle already detectable now with COBE or in the near
future with MAP can exist without a significant backreaction problem
provided the characteristic scale $k_{\rm c}\lesssim 10^{-2.5}M_{\rm
Pl}$ (but larger than the Hubble scale of inflation).  Similarly, a
modification of $1\%$ in principle observable by the Planck satellite
mission can be obtained if $s\gtrsim 2$.  Interestingly, these domains
encompass the Grand Unification scale $\sim10^{16}\,$GeV and possibly
the string scale.

  At this stage one should note that the issue of backreaction in the
present Universe~\cite{starobinsky}, which can be interpreted as the
production of gravitons of super-Planck momentum, applies to those
dispersion relations which violate WKB today, {\it i.e.} those for
which $\omega<a_0H_0$ today (notwithstanding singularities in the
dispersion relation). Since the comoving Hubble scale today is orders
of magnitude below the comoving Hubble scale of inflation, a
dispersion relation which broke the WKB approximation during inflation
does not necessarily imply production of gravitons today. This holds
in particular for those dispersion relations above which entail a
modification of the power spectrum without a strong backreaction
problem at the time of inflation.

  Finally one should note that the above situation may be encountered
for a wider class of dispersion relations, if the early time evolution
of the scale factor is different from inflationary expansion. Notably
consider a spacetime which is asymptotically Minkowski as
$\eta\to-\infty$, with a scale factor evolving as $a(\eta)=a_{\rm i}+
a_{\rm inf}(\eta/\eta_{\rm inf})^\beta$. Provided ${\rm d}\ln
\omega_{\rm phys}/{\rm d}\ln k_{\rm phys}$ is not singular, the WKB
approximation is valid asymptotically. One could reproduce the above
calculation with the substitution for the new scale factor, and the
final expression for the power spectrum would be the same [at late
times, $a_{\rm i}$ becomes negligible compared to the expansion term
in $a(\eta)$]. However the ratio $\eta_2/\eta_1$ then depends on $k$
if $\eta_2$ is in the far past when the term $a_{\rm i}$ cannot be
neglected even if $\beta =-1$.

\section{Conclusions}
\label{sec_concl}

  We have derived the stress-energy tensor for a free scalar field
with a general modified dispersion relation. In particular, we have
obtained the expression of the energy density and pressure in a FLRW
background. This result generalises previous
studies~\cite{jacobson,jacobson2} which were restricted to the
Corley-Jacobson dispersion relation. We have applied our calculation
of the stress-energy tensor to a series of examples to discuss the
equation of state of trans-Planckian modes and the issue of
backreaction in inflationary cosmology.

  We have first examined in details the possibility that the energy
contained in trans-Planckian modes with a frequency much smaller than
the Hubble expansion rate could account for the form of vacuum energy
density measured in the low-redshift Universe~\cite{mbk}. In the case
of dispersion relations with ultralow frequencies at high momenta, as
proposed in Ref.~\cite{mbk}, one cannot select a vacuum initial state
without ambiguity. Nevertheless, using well-motivated proposals for
this initial state, we have shown that the equation of state of these
trans-Planckian modes does not have the correct form, contrary to the
claim made in Ref.~\cite{mbk}. Moreover, the numerical value of the
energy density is not of the order of the critical energy density
unless fine-tuning is required, and we thus conclude that the scenario
proposed in Ref.~\cite{mbk} does not stand up to scrutiny.

  We have also discussed the issue of backreaction in trans-Planckian
inflationary cosmology. In particular, we have focused on a class of
dispersion relations for which $\omega_{\rm phys}< H$ for a finite
time interval for trans-Planckian comoving momenta. This is the
general class of dispersion relations for which the WKB approximation
is broken at some point during inflation, but is restored in the far
past as $\eta\to-\infty$. If the evolution is adiabatic all throughout
inflation up to horizon exit, it is known that the power spectrum is
not modified. In the above case, the WKB approximation is precisely
broken in the region where $\omega_{\rm phys}< H$, but is valid at
earlier and at later times.  We have obtained the analytical
expression for the amount of energy density stored in modes at late
time, and find that it is in general much larger than the background
energy density. We have computed the power spectrum of metric
fluctuations and showed that this power spectrum is not modified
except for its overall amplitude, independently of the dispersion
relation and whether WKB holds or not, if the inflationary period is
strictly de Sitter. This supports the belief that inflation is robust
to a change in the dispersion relation. If however the inflationary
period is not strictly de Sitter, then the power spectrum is tilted
with respect to the standard case (unmodified dispersion relation) and
superimposed oscillations appear. Finally we have exhibited a class of
dispersion relations for which the power spectrum of metric
fluctuations is strongly modified and the initial conditions can be
set up properly, since they are fixed in a region where the WKB
approximation holds.

  Our work thus completes the arguments developed in previous
works~\cite{mb,bm,niemeyer,np} on the relation between adiabaticity of
the mode evolution, the modification of the predictions of inflation,
and the magnitude of backreaction. In particular the following picture
seems to emerge: if the evolution of the modes is adiabatic all
throughout inflation up to horizon exit, then the power spectrum is
unmodified (or weakly modified), and backreaction is not an issue. If
however adiabaticity is broken at some point, then modifications to
the power spectrum are likely to appear (except if the background
spacetime is very close to de Sitter), but effects of backreaction
then appear generic, and one must use a higher order expansion of the
Einstein equations to derive meaningful conclusions. Finally, there
exists dispersion relations such that the backreaction is weak but
modifications to the power spectrum are not negligible, notably when
the ratio $k_{\rm c}/H_{\rm inf}$ is not too large, when the time
interval in which adiabaticity is broken is small, and when the
background spacetime is not de Sitter. However this obviously requires
fine-tuning, and overall a scale invariant power spectrum does indeed
appear robust against changes in the dispersion relation if
backreaction can be neglected.

\section*{Acknowledgements}
We thank Ted Jacobson for useful discussions and Robert Brandenberger
for comments; J.~P.~U thanks Renaud Parentani for commenting
Ref.~\cite{np} and I.~A.~P. for hospitality while this work was
carried out; L.~M. thanks F.~N.~R.~S. and C.~N.~R.~S. for financial
support and I.~A.~P.  for warm hospitality.


\appendix
\section{Notations}\label{AppA}

This appendix provides some identities satisfied by the vector field
$u_\mu$, the projector $\perp_{\mu\nu}$ and the derivative ${\cal
D}_\alpha$ that are implicitly used in the calculations of
Section~\ref{sec_2}. Since the norm of $u_\mu$ is conserved, one has
trivially $u_\alpha\nabla_\mu u^\alpha=0$. The covariant derivative of
$u_\mu$ can be conveniently decomposed as
\begin{equation}\label{dec_u}
\nabla_\mu u_\nu=\frac{1}{3}\theta\perp_{\mu\nu}
+\sigma_{(\mu\nu)}+\omega_{[\mu\nu]}-a_\mu u_\nu,
\end{equation}
where $A_{(\mu\nu)}\equiv(A_{\mu\nu}+A_{\nu\mu})/2$ and
$A_{[\mu\nu]}\equiv(A_{\mu\nu}-A_{\nu\mu})/2$ are the symmetrised and
antisymmetrised parts of $A_{\mu\nu}$ respectively. The shear
$\sigma_{\mu\nu}$ and vorticity $\omega_{[\mu\nu]}$ are tracefree and
purely spatial, i.e.
\begin{equation}
\sigma_{\mu\nu}g^{\mu\nu}=0,\qquad \sigma_{\mu\nu}u^\mu=0,\qquad
\omega_{\mu\nu}g^{\mu\nu}=0,\qquad \omega_{\mu\nu}u^\mu=0,
\end{equation}
and $a_\mu\equiv u^\nu\nabla_\nu u_\mu$ is the acceleration of the
observer comoving with $u_\mu$. The quantity $a_\mu$ is spatial, i.e. $a_\mu
u^\mu=0$ and one has $a_\mu=0$ for a geodesic. Using these relations,
the expansion rate $\theta$ can be written as
\begin{equation}
\theta = \perp^{\mu\nu}\nabla_\mu u_\nu.
\end{equation}
The integrability condition of the hypersurface $\Sigma$ implies that
$\omega_{\mu\nu}=0$ which is indeed satisfied if the definition 
of $u^{\mu }$ given in the main text is 
imposed. The projection operator $\perp_{\mu\nu}$ verifies
\begin{equation}
\perp_\mu^\mu=3,\qquad \perp_\mu^\nu u^\mu=0,\qquad
\perp_\alpha^\beta\perp_\beta^\gamma=\perp_\alpha^\gamma.
\end{equation}
For the FLRW case, it follows that $\theta=3H$, $a_\mu=
\sigma_{\mu\nu}=\omega_{\mu\nu}=0$.
\par
Finally, if we denote by $\sigma^i$ the internal coordinates on the
hypersurface $\Sigma$ defined by the embedding $x^\mu=\bar
x^\mu(\sigma^i)$, then the three dimensional spatial metric is given
by $\perp_{ij}=g_{\mu\nu} \partial _i\bar{x}^\mu \partial _j\bar{x}^\nu $ and it
follows that $\DD^2\phi$ is given directly by
\begin{equation}\label{A13}
{\cal D}^2\phi=\frac{1}{\sqrt{\perp}}\partial_i(\sqrt{\perp}\partial^i\phi),
\end{equation}
where $\perp$ is the determinant of $\perp_{ij}$.
Hence ${\cal D}_\mu{\cal D}^\mu\phi$ is the three dimensional Laplacian
as defined by the observer comoving with $u^\mu$. With the decomposition 
Eq.~(\ref{dec_u}), it is easily shown using
the identity $
u^\alpha
u^\beta\nabla_\alpha\nabla_\beta\phi=\ddot\phi-a^\alpha\nabla_\alpha\phi
$ that
\begin{equation}\label{lapl}
{\cal D}^2\phi=\Box\phi+\ddot\phi-a^\alpha\nabla_\alpha\phi+\theta\dot\phi,
\end{equation}
where $\Box\equiv\nabla_\mu\nabla^\mu$ denotes the four dimensional
d'Alembertian.

\section{Derivation of the stress--energy
tensor in the particular case of the Corley-Jacobson dispersion 
relation}
\label{app_B}

We detail in this appendix the simple case where only $b_{11}$ does
not vanish in the corrective Lagrangian. This Lagrangian contains
derivatives of the metric that must be varied to obtain the
stress-energy tensor. With $\delta\sqrt{-g}/\delta
g^{\mu\nu}=-\sqrt{-g}g_{\mu\nu}/2$, this stress-energy tensor can be
written as
\begin{equation}\label{tmunu}
T^{\mu\nu}=-2g^\mu_\alpha g^\nu_\beta
\frac{\delta{\cal L}}{\delta g^{\alpha\beta}}+{\cal
L}g^{\mu\nu}-\frac{2}{\sqrt{-g}}\partial_\rho\left(\sqrt{-g}
\frac{\delta{\cal L}}{\delta\partial_\rho g_{\mu\nu}}\right).
\end{equation}
For the present case ${\cal L}_{_{\rm cor}}=-b_{11}(\DD^2\phi)^2$, it
leads to
\begin{eqnarray}
T^{\mu\nu}&=&\partial^\mu\phi\partial^\nu\phi-\frac{1}{2}\left(\partial_\alpha
\phi\partial^\alpha\phi\right)g^{\mu\nu}+2\lambda u^\mu u^\nu
-b_{11}\left(\DD^2\phi\right)^2g^{\mu\nu}
\nonumber\\
&&\qquad+b_{11}\left\{2g^\mu_\alpha g^\nu_\beta\frac{\delta
\left(\DD^2\phi\right)^2}{\delta g^{\alpha\beta}}
+\frac{2}{\sqrt{-g}}\partial_\rho\left[\sqrt{-g}
\frac{\delta\left(\DD^2\phi\right)^2}{\delta\partial_\rho
g_{\mu\nu}}\right]\right\}.
\end{eqnarray}
The first term in brackets can be written, using Eq.~(\ref{defd}),
\begin{equation}\label{d1}
\frac{\delta\left(\DD^2\phi\right)}{\delta
g^{\mu\nu}}=\nabla_{(\mu}\nabla_{\nu)}\phi+\dot\phi
\nabla_{(\mu}u_{\nu)}+\theta u_{(\mu}\nabla_{\nu)}\phi+ 2u^\alpha
u_{(\mu}\nabla_{\nu)}\nabla_\alpha\phi-\perp^{\alpha\beta}
\Gamma_{\alpha\beta(\mu}\nabla_{\nu)}\phi-\dot\phi g^{\alpha\beta}
\Gamma_{\alpha\beta(\mu}u_{\nu)},
\end{equation}
while the second term gives
\begin{equation}\label{d2}
\frac{\delta\left(\DD^2\phi\right)}{\delta\partial_\rho g_{\mu\nu}}=
-\frac{1}{2}\left[2\perp^{\rho(\mu}\nabla^{\nu)}\phi-\perp^{\mu\nu}
\nabla^\rho\phi+2\dot\phi g^{\rho(\mu}u^{\nu)}-\dot\phi g^{\mu\nu}
u^\rho\right].
\end{equation}
The derivative of this equation finally reads
\begin{eqnarray}\label{d3}
\frac{1}{\sqrt{-g}}\partial_\rho\left[\sqrt{-g}
\frac{\delta\left(\DD^2\phi\right)}{\delta\partial_\rho
g_{\mu\nu}}\right]&=&-\frac{1}{2}\nabla_\rho\left[\perp^{\rho(\mu}
\nabla^{\nu)}\phi-\perp^{\mu\nu}
\nabla^\rho\phi+\dot\phi g^{\rho(\mu}u^{\nu)}-\dot\phi g^{\mu\nu}
u^\rho\right]\nonumber\\
&&+\perp^{\rho\alpha}\Gamma_{\rho\alpha}^{(\mu}\nabla^{\nu)}\phi
+\dot\phi g^{\rho\alpha}\Gamma_{\rho\alpha}^{(\mu}u^{\nu)}.
\end{eqnarray}
It can be checked that the terms involving Christoffel symbols in
Eqs.~(\ref{d1}) and (\ref{d2}) strictly vanish, a necessary condition to
the covariance of the stress-energy tensor. Finally, we end up with
\begin{eqnarray}
T_{\mu\nu}&=&\partial_\mu\phi\partial_\nu\phi-\frac{1}{2}\left(\partial_\alpha
\phi\partial^\alpha\phi\right)g_{\mu\nu}+2\lambda u_\mu u_\nu
+T_{\mu \nu}^{\rm (F)}
\nonumber\\
&&+b_{11}\left({\cal D}^2\phi\right)\biggl[2\Box\phi\perp_{\mu\nu}
+2\left(\ddot\phi+\theta\dot\phi\right)g_{\mu\nu}-\left({\cal
D}^2\phi\right)g_{\mu\nu}\nonumber\\
&&+4u_{(\mu}\nabla^\alpha u_{\nu)}\nabla_\alpha
\phi-4u_{(\mu}\nabla_{\nu)}u^\alpha\nabla_\alpha\phi
-4a_{(\mu}\nabla_{\nu)}\phi\biggr]\nonumber\\
&&-2b_{11}\biggl[2\nabla_{(\mu}\phi\nabla_{\nu)}\left({\cal
D}^2\phi\right)+2\left({\cal D}^2\phi\right)^. u_{(\mu}
\nabla_{\nu)}\phi -\perp_{\mu\nu}\nabla_\alpha\phi
\nabla^\alpha\left({\cal D}^2\phi\right)\nonumber\\
&&+2\dot\phi
u_{(\mu}\nabla_{\nu)}\left({\cal D}^2\phi\right)-
\dot\phi\left({\cal D}^2\phi\right)^.g_{\mu\nu}\biggr].
\end{eqnarray}
\par
The expression of the Lagrange multiplier $\lambda$ results from the
equation of motion for $u^\mu$
\begin{equation}\label{ml}
u^\mu\frac{\delta{\cal L}}{\delta
u^\mu}=\frac{1}{\sqrt{-g}}u^\mu\partial_\nu\left(\sqrt{-g}\frac{\delta{\cal
L}}{\delta \partial_\nu u_\mu}\right),
\end{equation}
which gives
\begin{equation}\label{temp}
\lambda=b_{11}\left[\left(\DD^2\phi\right)\left(\ddot\phi+
\theta\dot\phi-2a^\alpha\nabla_\alpha\phi\right)
-\left(\DD^2\phi\right)^.\dot\phi\right]+\lambda ^{\rm (F)}.
\end{equation}
It follows that the energy density and the pressure measured
by the observer comoving with $u^\mu$ (i.e. $\rho=T_{\mu\nu}u^\mu
u^\nu$ and $p=T_{\mu\nu}\perp^{\mu\nu}/3$) are respectively given by
\begin{eqnarray}
\rho &=& \dot\phi^2+\frac{1}{2}\nabla_\alpha\phi
\nabla^\alpha\phi+b_{11}\left(\DD^2\phi\right)^2
+u^{\mu }u^{\nu }T_{\mu \nu}^{\rm (F)}+2\lambda ^{\rm (F)},
\\
\label{41}
p &=& \frac{1}{3}\dot\phi^2-\frac{1}{6}\nabla_\alpha\phi
\nabla^\alpha\phi+\frac{b_{11}}{3}\left(\DD^2\phi\right)
\left[3\left(\DD^2\phi\right)+2a^\alpha\nabla_\alpha\phi\right]
+\frac{2b_{11}}{3}\dot\phi\left(\DD^2\phi\right)^.\nonumber\\
&&+\frac{2b_{11}}{3} \nabla_\alpha\phi\nabla^\alpha(\DD^2\phi)
-\frac{1}{3}\perp ^{\mu \nu}T_{\mu \nu}^{\rm (F)}.
\end{eqnarray}
In the case of the Minkowski metric, the mean values of the 
previous expressions calculated in a thermal state reduce to 
the formulas obtained in Refs.~\cite{jacobson,jacobson2}. 
\par 
Finally, the field equation for $\phi$ is obtained by varying the
action to obtain
\begin{equation}
\frac{\delta{\cal
L}\sqrt{-g}}{\delta\phi}=\partial_\mu\left(\frac{\delta{\cal
L}\sqrt{-g}}{\delta\partial_\mu\phi}\right)-\partial_{\mu\nu}
\left(\frac{\delta{\cal
L}\sqrt{-g}}{\delta\partial_{\mu\nu}\phi}\right).
\end{equation}
hence
\begin{equation}
\Box\phi=-2b_{11}\left[\DD^2\left(\DD^2\phi\right)
+2a^\nu\nabla_\nu\left(\DD^2\phi\right)+\nabla_\mu
a^\mu\left(\DD^2\phi\right)\right].
\end{equation}
It can be checked that the general formulae given Section~\ref{sec_2}, when
applied to the special case of the present Corley-Jacobson dispersion
relation, reduce to the equations given in this Appendix.

\end{document}